\documentclass[pra,twocolumn,showpacs,preprintnumbers,amsmath,amssymb]{revtex4}

\usepackage{graphicx}
\usepackage{dcolumn}
\usepackage{bm}

\begin{document}

\title{Density modulations in an elongated Bose-Einstein condensate released from a disordered potential}

\author{D.~Cl\'ement}
\author{P.~Bouyer}
\author{A.~Aspect}
\author{L.~Sanchez-Palencia}
\affiliation{Laboratoire Charles Fabry de l'Institut d'Optique, CNRS
and Univ. Paris-Sud, Campus Polytechnique, RD 128, F-91127 Palaiseau
cedex, France}
\date{\today}

\begin{abstract}
We observe large density modulations in time-of-flight images of elongated
Bose-Einstein condensates, initially confined in a harmonic trap and
in the presence of weak disorder.
The development of these modulations during the time-of-flight and their
dependence with the disorder are investigated.
We render an account of this effect using numerical and analytical calculations.
We conclude that the observed large density modulations originate from the weak
initial density modulations induced by the disorder, and not from
initial phase fluctuations (thermal or quantum).
\end{abstract}


\maketitle

Gaseous Bose-Einstein condensates (BEC) in disordered potentials
\cite{LyePRL2005,ClementPRL2005,FortPRL2005,SchultePRL2005,ClementNJP2006}
offer controllable systems to study
open basic questions on the effects of disorder in quantum media
\cite{ColdDisoTh}.
In this respect, a
still debated question relies on the nature of the
disorder-induced superfluid-insulator transitions \cite{phillips2003}
which can originate from strong fluctuations of either the density or the phase.
This question can be addressed experimentally with gaseous BECs in optical disorder \cite{LyePRL2005,ClementPRL2005,FortPRL2005,SchultePRL2005,ClementNJP2006} as
both density and phase can be measured directly \cite{anglin2002}.
In addition to its fundamental interest, this study is important for BECs on chips \cite{AtomChips},
and can also shed some light on the physics of dirty superconductors
\cite{desoSC} and granular metals \cite{beloborodov2007}.

\begin{figure}[b!]
\begin{center}
\includegraphics[width=8.cm]{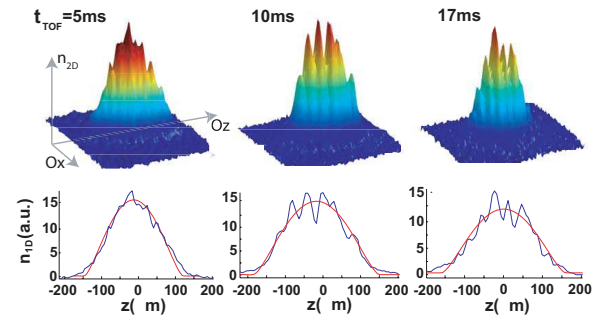}
\end{center}
\caption{
Upper panel: TOF images of an expanding disordered BEC for
three different times of flight. The vertical axis represents the column density
along the $y$ axis.
Lower panel: axial 1D density profiles
$n_{1\textrm{D}}(z)$ (column density integrated along the $x$ axis; blue)
and 1D TF parabolic profiles
$n_{1\textrm{D}}^0(z)$ (red). The amplitude of the disorder
is $\gamma=V_\textrm{R}/\mu=0.41$.} \label{FigAbsorpImg}
\end{figure}

In the presence of repulsive interactions, Anderson localization is suppressed in a
stationary BEC, and weak disorder results in small density modulations \cite{SanchezPalenciaPRA2006}.
However, one may wonder whether weak disorder {\it may} significantly affect the coherence of
a connected BEC and entail large phase fluctuations.
It has been suggested \cite{LyePRL2005,LewensteinAdvPhys2007},
by analogy with elongated (non-disordered) quasi-BECs,
that the observation of large fringes in \textit{time-of-flight} (TOF)
images of disordered BECs
(see Fig.~\ref{FigAbsorpImg} and \cite{LyePRL2005,chen2007,hannover}) may 
signal
strong initial phase fluctuations \cite{PetrovPRL2001,DettmerPRL2001,RichardPRL2003}.
For quasi-BECs, such fringes in TOF images are {\it indeed} a signature of
initial phase fluctuations \cite{DettmerPRL2001}.
However, for disordered BECs, no systematic study of these fringes has been reported so far
and their relation to disorder-induced phase fluctuations is still unclear.

In this paper, we report a detailed study of the density modulations
in the TOF images of an elongated, non-fragmented
three-dimensional (3D) BEC initially placed in a weak 1D disordered potential.
Our main experimental result is that the fringes in the TOF images
are reproduced from shot to shot when using the {\it same} realization of
the disorder. This excludes disorder-induced
phase fluctuations (thermal or quantum) in the trapped BEC
as the origin of the fringes observed after TOF,
for the parameter range of the experiment (relevant also for 
\cite{LyePRL2005,chen2007,hannover}).
Using analytical and numerical calculations, which do not include initial phase fluctuations,
we show that the fringes actually develop during the TOF according to the following scenario.
Just after release, the initial weak density
modulations (induced by the disorder onto the trapped BEC) imprint a phase
with axial modulations and transversal invariance.
Then, the resulting axial phase modulations
are converted into large axial density modulations.

The experiment is detailed in Refs.~\cite{ClementPRL2005,ClementNJP2006}.
We form a
cigar-shaped BEC of $^{87}$Rb atoms in a Ioffe-Pritchard trap of
frequencies $\omega_z/2\pi \! \! = \! \! 6.7$Hz and
$\omega_{\bot}/2\pi \! \! = \! \!  660$Hz. The BEC atom number is
$N_0 \sim 3 \times 10^5$, the length $2 \! L_{\rm{TF}} \! \simeq \! 300 \mu$m, and the chemical potential
$\mu/2\pi\hbar \simeq 4.5$kHz. We create a 1D speckle (disordered) potential
along the $z$ axis.
The correlation function is $C(z)=V_\textrm{R}^2 \textrm{sinc}^2
(z/\sigma_\textrm{R})$ where both amplitude $V_\textrm{R}$ and correlation length
$\sigma_\textrm{R}$ down to $0.33 \mu$m can be controlled \cite{ClementPRL2005,ClementNJP2006}.
The results presented in this work correspond to $\sigma_\textrm{R} \simeq 1.7\mu$m
\cite{notesigma}.
In the experiment, we wait 300ms for the BEC to reach equilibrium in the presence of disorder
and then switch off abruptly both magnetic and speckle potentials.
We then take absorption images of the expanding cloud after a
variable time of flight $t_{\rm{TOF}}$ with typical images shown
in Fig.~\ref{FigAbsorpImg}.

These images show large density modulations along the axis $z$ of
the disorder. To measure their amplitude, we first extract a 1D
axial density $n_{1\textrm{D}}(z)$ by integrating the column
density over the second transverse direction $x$. We then define
$\eta(z)$ as the normalized deviations of the 1D density from the 1D
parabolic Thomas-Fermi (TF) profile $n_{1\textrm{D}}^0(z)$ which
fits best the 
data (red line in Fig.~\ref{FigAbsorpImg}), so that
$n_{1\textrm{D}}(z)=n_{1\textrm{D}}^0(z) [1+ \eta(z)]$. Finally we
calculate the standard deviation of $\eta(z)$ over a given length
$L$: $\Delta \eta = \sqrt{\frac{1}{L} \int_L \textrm{d}z \
\eta^2(z)}$ (here, $\int_L \textrm{d}z \ \eta(z) = 0$). The
calculation of $\Delta \eta$ is restricted to 70\% of the BEC total
length ($L=1.4 L_{\rm{TF}}$) to avoid the edges
where thermal atoms are present.
Two imperfections reduce
the measured $\Delta \eta$
compared to the real value. First, our imaging system has a finite
resolution $L_{\textrm{res}}=8.5\mu$m, larger than the variation
scale $\sigma_{\textrm{R}}$ of the disorder. This effect is
quantified by measuring the modulation transfer function (MTF) of the
imaging system. Second, a slight mis-alignment of the probe beam --
which is not exactly perpendicular to axis $z$ -- also reduces the
contrast of the fringes \cite{DettmerPRL2001}. This effect is more
difficult to quantify as angles smaller than our uncertainty on the
probe angle ($1^\circ$) can drastically reduce the
contrast. We find that numerics reproduce our
experimental results assuming a mis-alignment of $0.33^\circ$.

\begin{figure}[t!]
\begin{center}
\includegraphics[width=6.cm]{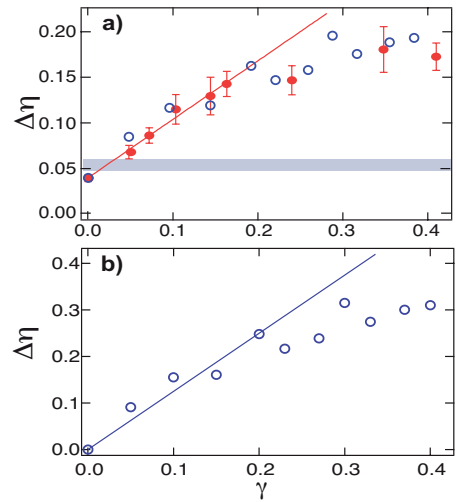}
\end{center}
\caption{\textbf{a)} Experimental results (red points)
for the density modulations at $\omega_{\bot}t_{\textrm{TOF}}= 62.2$
versus the amplitude of the disorder.
The shaded area corresponds to phase fluctuations in our initial
elongated BEC, as calculated in Ref.~\cite{DettmerPRL2001}
(the error bar reflects the uncertainty on the temperature).
\textbf{b)} Corresponding numerical results, taking into account the finite
resolution of our optics but not the misalignment of the probe beam.
The open blue circles in a) show the same data including an offset accounting
for the small initial phase fluctuations and the correction corresponding to
the probe angle.}
\label{FigPowerEvolDn}
\end{figure}

\begin{figure}[t!]
\begin{center}
\includegraphics[width=6.cm]{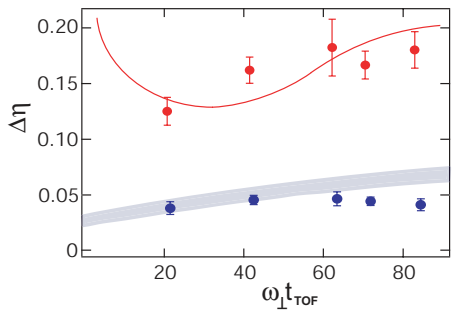}
\end{center}
\caption{ Time evolution of the measured density modulations $\Delta
\eta$ during a TOF for $\gamma=0$ (no disorder; blue points) and
$\gamma=0.41$ (with disorder, red points). The shaded area
corresponds to the calculation of Ref.~\cite{DettmerPRL2001} taking
into account the uncertainty on the temperature. The solid red line
is the result of numerical calculations for $\gamma=0.4$ (see
text).
} \label{FigTimeEvolDn}
\end{figure}

We first study the amplitude of the normalized density modulations,
$\Delta \eta$, as a function of the amplitude of the disorder $\gamma
= V_{\rm{R}}/\mu$ at a given time-of-flight $t_{\rm{TOF}}=15.3$ms
($\omega_\perp t_{\rm{TOF}}=62.2$) with experimental results plotted
in Fig.~\ref{FigPowerEvolDn}a. In the absence of disorder,
we observe non-vanishing density modulations ($\Delta\eta_0 \simeq
0.037$) larger than the noise in the background of the
images ($\Delta\eta_{\rm{n}} \simeq 0.015$).
They are interpreted as small but non-zero phase fluctuations initially present
in our elongated BEC \cite{PetrovPRL2001}.
The calculation of their contribution to the
density modulations in the TOF images as determined in Ref.~\cite{DettmerPRL2001}
agrees with our data (see Fig.~\ref{FigPowerEvolDn}a and Fig.~\ref{FigTimeEvolDn}).
In the presence of disorder, we find that, for small values of $\gamma$ (typically $\gamma<0.2$),
$\Delta \eta$ grows with $\gamma$ as $\Delta \eta=\Delta \eta_0 + 0.64(3) \gamma$.
For larger
values of $\gamma$, the disordered BEC is fragmented either in the
trap or during the expansion, and $\Delta \eta$ has a maximum
value of $\Delta\eta \simeq 0.17$ in the experiment
\cite{noteContrasteEvolution}.

We also perform numerical integrations of the 3D Gross-Pitaevskii
equation (GPE) for the expanding disordered BEC and extract
$\Delta\eta$
as in the experiments.
The numerics do not include any initial phase
fluctuations. We find a linear dependence of $\Delta \eta$ versus
$\gamma$, $\Delta\eta \simeq 3.5\gamma$ for the bare numerical
results, and $\Delta \eta \simeq 1.23\gamma$ if we take into account
the finite resolution of the imaging system but not the
mis-alignment of the probe (see Fig.~\ref{FigPowerEvolDn}b). In
fact, we find that the numerics agree with the
experiments
if, in addition to an offset $\Delta\eta_0$ to
mimic the small initial phase fluctuations, we include the
systematic correction associated to a probe angle of $0.33^\circ$
(see Fig.~\ref{FigPowerEvolDn}a).

We now examine the TOF dynamics of the disordered BEC and plot
$\Delta \eta$ versus $t_{\rm{TOF}}$ in Fig.~\ref{FigTimeEvolDn}. In
the presence of disorder, the observed density modulations (red
points) are clearly enhanced compared to those in the absence of
disorder (blue points). We also observe that the density modulations
first develop, and then saturate.
The dynamics of their development
is reproduced well by our numerical calculations (solid red line),
if we take into account all imperfections of our imaging system.

After correcting all experimental imperfections, the density modulations
in the TOF images turn out to
be larger ($\Delta\eta \simeq 3.5 \gamma$) than the ones of
the trapped BEC before TOF ($\Delta\eta=2 \gamma$) \cite{note2}. One
may wonder whether these large density modulations in the TOF images
reveal phase fluctuations induced by the disorder in the initial BEC
\cite{LyePRL2005,LewensteinAdvPhys2007}. Actually, several
arguments lead us to conclude that it is not so. First, our
numerics, which reproduce
the experimental data,
do {\it not} include initial phase fluctuations. Second, the numerical
diagonalization of the Bogolyubov equations indicates that the
disorder hardly affect the excitation spectrum of a 1D BEC for
the experimental parameters \cite{notebogo}. Last but
not least, we have observed identical density modulations in
successive experiments performed with the {\it same} realization of
the disorder
(see also Refs.~\cite{chen2007,hannover}).
Hence, averaging
over various images taken with the same realization of the disorder does
not wash out the fringes \cite{noteaverage} and this
excludes initial random fluctuations, quantum or thermal \cite{notesigma}.

We now develop an analytical model for the evolution of the
BEC density profile during the TOF,
which shows explicitly how a weak disorder
leads to large density modulations after a long-enough TOF,
{\it without initial phase fluctuations}.
Although the probability of fragmentation is small
for weak-enough disorder, it may happen that the BEC is fragmented into a
small number of fragments. However, for the considered
$t_\textrm{TOF} \lesssim 1/\omega_z$,
the fragments will only weakly overlap as the axial expansion
is small. Therefore, we neglect fragmentation in our model.

In the absence of disorder,
the TOF expansion of a BEC initially trapped in a harmonic potential
in the Thomas-Fermi regime is self-similar \cite{KaganPRA1996},
so that
\begin{equation}
\psi(\vec{r},t)=\Big[ {\prod }_j b_j(t) \Big]^{-1/2} \phi \left ( \{ x_j/b_j(t)
\},t \right ) e^{i \theta_0(\vec{r},t)},
\label{scaling}
\end{equation}
with $j \! = \! 1,..,3$ the spatial directions,
$\theta_0(\vec{r},t) = (m/2 \hbar) \sum_j (\dot{b}_j/b_j) x_j^2$
the dynamical phase
and
$\phi$ the (time-independent)
wavefunction of the BEC in the trap.
The scaling factors $b_j(t)$ are governed by the equations
$\ddot{b}_j = \omega_j^2 /(b_j \prod_k b_k)$ with the initial
conditions $b_j(0)=1$ and $\dot{b}_j(0)=0$ \cite{KaganPRA1996}.
In the presence of
disorder, we use the scaling~(\ref{scaling}) and we write the (now
time-dependent) wavefunction $\phi(\rho,z,t) \! = \!
\sqrt{\tilde{n}(\rho,z,t)} e^{i\tilde{\theta}(\rho,z,t)}$ where
$\rho \! = \! \sqrt{x^2 \! + \! y^2}$ is the radial coordinate. In
the absence of phase fluctuations, $\phi(\rho,z,t \! = \! 0)$ is
real (up to a homogeneous phase) as it is the ground state of the
trapped, disordered BEC. The TOF dynamics is then governed by two
coupled equations for the density $\tilde{n}$ and the phase
$\tilde{\theta}$ which are equivalent to the complete time-dependent
GPE.

Let us introduce now a couple of approximations.
First, in elongated 3D BECs, the expansion for $\omega_z t \lesssim 1$
is mainly radial and $b_z(t) \simeq 1$.
Second, we assume not too large perturbation of the density, so that
$\tilde{n} = \tilde{n}_0 + \delta \tilde{n}$ with
$\tilde{n}_0$, the density in the absence of disorder and
$\delta \tilde{n} \ll \tilde{n}_0$. Using local density approximation,
we can neglect all the spatial derivatives of $\tilde{n}_0$.
We also neglect the radial derivatives of $\delta\tilde{n}$ since the
1D disorder induces short-range spatial inhomogeneities
mainly along the z-axis. We are thus left with the equations
\begin{eqnarray}
\partial_t \delta \tilde{n} &=& - (\hbar/m) \tilde{n}_0 \partial^2_z \tilde{\theta} \label{Eq:CoupledDnPhase1} \\
- \hbar \partial_t \tilde{\theta} &=& g \delta \tilde{n}
/b_{\bot}^2 + (\hbar^2/2 m)\ [|\partial_{z} \tilde{\theta}|^2 -
\partial^2_z \delta \tilde{n}/2 \tilde{n}_0 ].
\label{Eq:CoupledDnPhase2}
\end{eqnarray}

In a first stage, the initial small inhomogeneities of the density
induced by the 1D disordered potential before TOF, $\delta\tilde{n} \simeq -V(z)/g$
\cite{SanchezPalenciaPRA2006}, hardly evolve since $\partial_t
\delta\tilde{n} (t=0) = 0$. They are however crucial as they act as
an inhomogeneous potential which induces the development of a phase
modulation $\tilde{\theta}(z,t)$ at the beginning of the TOF
\cite{noteimprinting}.
From Eqs.~(\ref{Eq:CoupledDnPhase1}),(\ref{Eq:CoupledDnPhase2}), we
find
\begin{eqnarray}
& & \tilde{\theta}(z,t) \simeq \arctan(\omega_{\bot} t) [V(z)/\hbar \omega_{\bot}]
\label{Eq:SolutionTheta} \\
\textrm{and} & & \delta\tilde{n} \simeq -V(z)/g
-(\tilde{n}_0 \partial_z^2 V(z) / m\omega_\perp^2) F(\omega_\perp t)
\label{Eq:SolutionN}
\end{eqnarray}
where $F(\tau) \!\! = \!\! \int_0^\tau \textrm{d}\tau'
\arctan(\tau') \!\! = \!\! \tau \arctan(\tau) \!\! - \!\! \ln
\sqrt{1+\tau^2}$. From Eq.~(\ref{Eq:SolutionN}), we then find
\begin{equation}
\Delta\eta (t) \simeq 2 \left( \frac{V_\textrm{R}}{\mu} \right)
             \left[1 - \frac{2}{3} \left(\frac{\mu}{\hbar\omega_\perp}\right)^2  \!\!
         \left(\frac{\xi}{\sigma_\textrm{R}}\right)^2  \!\! F (\omega_\perp t) \right].
\label{Eq:etashort}
\end{equation}
Hence, $\Delta \eta$ first slightly decreases. It is easily
understood as an interacting BEC initially at rest will tend to fill
its holes when released from the disordered potential. The
solution~(\ref{Eq:SolutionTheta}),(\ref{Eq:SolutionN}) is valid as
long as the contribution of the last two terms in
Eq.~(\ref{Eq:CoupledDnPhase2}) remains small [{\it i.e.} for
$\omega_\perp t \ll (\sigma_\textrm{R}/\xi)^2$ and $\omega_\perp t
\ll (\sigma_\textrm{R}/\xi) (\hbar \omega_\perp /
\sqrt{V_\textrm{R}\mu})$]. In addition, it requires that the density
modulations do not vary much ({\it i.e.} the second rhs term in
Eq.~(\ref{Eq:SolutionN}) is small compared to the first one). For
the experimental parameters, the last condition is the most
restrictive. It defines a typical time $t_0 = (1/\omega_\perp)
F^{-1} \left[
             (\sigma_\textrm{R}/\xi)^2 (\hbar\omega_\perp/\mu)^2/2
 \right]$
during which the radial expansion imprints a phase modulation due to
the initial inhomogeneities
of the BEC density
created by the disorder before the TOF.
In particular, if $t_0 \gg 1/\omega_\perp$, the phase modulations
freeze at $\tilde{\theta}(z) \simeq (\pi/2)[V(z)/\hbar \omega_{\bot}]$.

\begin{figure*}[t!]
\begin{center}
\includegraphics[width=15.cm]{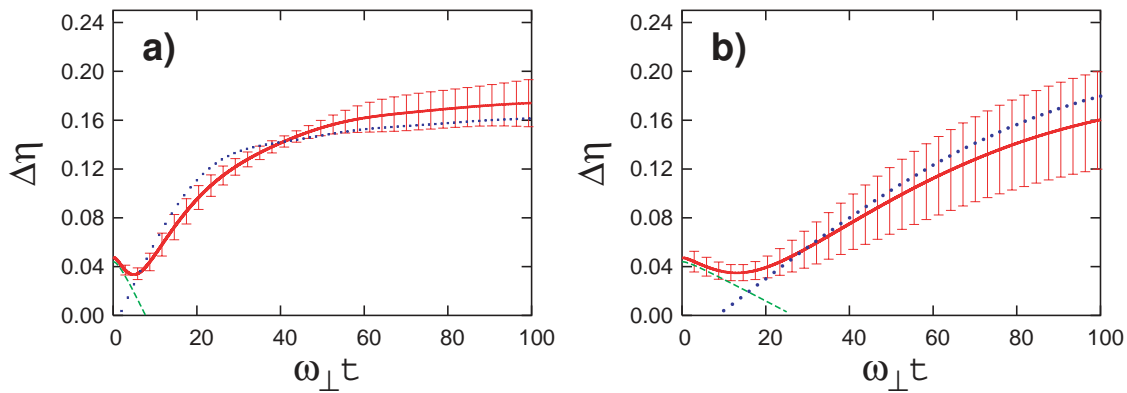}
\end{center}
\caption{
Dynamics of density modulations
as obtained numerically
(solid red line with error bars)
and comparisons with
Eq.~(\ref{Eq:etashort}) (green dashed line)
and Eq.~(\ref{Eq:denslongsolbis}) (blue dotted line).
{\bf a)} The parameters are the same as
in the experiment (in particular, $\sigma_\textrm{R}=1.7\mu$m)
with $V_\textrm{R}=0.02\mu$.
Here $\omega_\perp t_0 \simeq 3$ and we have used $t_1=0.5 t_0$.
{\bf b)} Same as a) but with $\sigma_\textrm{R}=3.4\mu$m.
Here, $\omega_\perp t_0 \simeq 8.5 \gg 1$, so that we have used $t_1=t_0$ (see text).
}
\label{fluct}
\end{figure*}

In a second stage, the phase modulations are converted into density
modulations similarly as thermal phase fluctuations do during the
TOF of an elongated quasi-BEC \cite{DettmerPRL2001}. For $t \gtrsim
t_1$ where $t_1$ is a typical time much longer than $1/\omega_\perp$
(see below), the scaling parameter $b_\perp (t)$ becomes large so
that the first rhs term in Eq.~(\ref{Eq:CoupledDnPhase2}) can now be
neglected. Assuming small phase gradients, we are left with the
equation $\partial_t^2 \delta \tilde{n}_k + \hbar^2 k^4 \delta
\tilde{n}_k /4m^2=0$ where $\delta \tilde{n}_k (\rho,t)$ is the 1D
Fourier transform of $\delta\tilde{n}$ along $z$, and whose solution
reads $\delta \tilde{n}_k (t) = \delta \tilde{n}_k (t_1) \cos\left[
(\hbar k^2/2m) (t-t_1) \right] + (2m \delta \dot{\tilde{n}}_k
(t_1)/\hbar k^2) \sin\left[ (\hbar k^2/2m) (t-t_1) \right]$. If $t_0
\gg 1/\omega_\perp$, we can take $1/\omega_\perp \ll t_1 \leq t_0$
and the exact value of $t_1$ does not matter much (we use $t_1 \! =
\! t_0$). If $t_0 \lesssim 1/\omega_\perp$, the determination of
$t_1$ is not straightforward but can be found through fitting
procedures, for instance. Then, according to
Eq.~(\ref{Eq:SolutionN}), $\delta \tilde{n}_k (t_1) \simeq -V(z)/g$
is mainly determined by the initial density modulations of the
trapped BEC while $\delta \dot{\tilde{n}}_k (t_1) \simeq
-(\tilde{n}_0 \partial_z^2 V(z) / m\omega_\perp)
\arctan(\omega_\perp t_1)$ results from the phase modulations
created in the first stage of the TOF. For $\hbar\omega_\perp \ll
\mu$ as in the experiment, the cosine term can be neglected and we
find
\begin{equation}
\Delta \eta (t) \simeq \sqrt{8}
(V_\textrm{R}/\hbar \omega_{\bot})
\arctan (\omega_{\bot} t_1) I[\sigma_\textrm{R},t-t_1]
\label{Eq:denslongsolbis}
\end{equation}
where $I(\sigma_\textrm{R},t) = \sqrt{\int_0^1 \textrm{d}\kappa\
(1-\kappa) \sin^2\left[ (2\hbar t/m\sigma_\textrm{R}^2) \kappa^2 \right]}$
for a speckle potential.

Numerical integrations of the complete 3D GPE confirm the expected
behavior of $\Delta \eta$ during the TOF at short [Eq.~(\ref{Eq:etashort})]
and long [Eq.~(\ref{Eq:denslongsolbis})] times as shown in Fig.~\ref{fluct}.
This validates our scenario in a quantitative manner.

Three remarks are in order.
First, we find that,
due to the development of phase modulations in the first stage of the TOF,
the density modulations in the expanded BEC
($\Delta \eta \propto V_\textrm{R}/\hbar \omega_{\bot}$)
can be larger than those
in the trapped BEC ($\Delta \eta \propto V_\textrm{R}/\mu$)
if $\mu > \hbar\omega_\perp$.
Second, the density pattern is completely determined by the
realization of the disorder.
Third, Eq.~(\ref{Eq:denslongsolbis}) shows that
the density modulations saturate at
$\Delta \eta \simeq \sqrt{2} (V_\textrm{R}/\hbar \omega_{\bot})
\arctan (\omega_{\bot} t_1)$ for very long times $t$ \cite{notelongtimes}.
These properties are in qualitative agreement with the
experimental observations.

In conclusion, we have shown that the large fringes observed in this
work and Refs.~\cite{LyePRL2005,chen2007,hannover} in TOF images  of disordered BECs
{\it do not} rely on initial disorder-induced phase fluctuations.
They actually result from the TOF process of a
BEC with small initial density modulations but without phase fluctuations.
A phase modulation determined by the weak initial modulations of the BEC density first
develops and is later converted into large density modulations.
Our analytical calculations based on this scenario agree with numerical calculations
and experimental observations.
Nevertheless, our results do not exclude that, in different regimes, disorder might enhance
phase fluctuations. This question is crucial in connection to the nature of superfluid-insulator
transitions in the presence of disorder.
Dilute BECs can help answering it as their phase coherence
can be probed with accuracy \cite{DettmerPRL2001,RichardPRL2003}.
Revealing this possible effect via TOF images requires taking into
account the phase modulation which develops during the first stage of the TOF
as demonstrated here.

We thank P. Chavel for useful discussions and J. Retter and
A. Var´on for their help in the early stage of the experiment.
This work was funded by the French DGA, MENRT, ANR and IFRAF
and by the ESF program QUDEDIS.

\bibliography{apssamp}

\begin{thebibliography}{99}
\bibitem{notesigma}
We found similar experimental results for $\sigma_\textrm{R}=0.33\mu$m.
\bibitem{noteContrasteEvolution}
For stronger disorder ($\gamma > 0.5$), quantum fluctuations of the
phases of the BEC fragments may reduce the contrast of the
fringes in TOF images.
\bibitem{noteaverage}
The pictures of Fig.~\ref{FigAbsorpImg} are actually averaged over
six TOF images with the same disordered potential.
\bibitem{notelongtimes}
This is valid for $t < 1/\omega_z$. However, at even longer times, the axial
expansion of the BEC may reduce $\Delta\eta$.
\end{thebibliography}

\end{document}